\documentclass[9pt,twocolumn,twoside]{osajnl}

\journal{ol} 

\setboolean{shortarticle}{true} 

\title{Controlled generation of ultrafast vector vortex beams from a mode-locked fiber laser}

\author[1]{Kun Huang}
\author[1]{Jing Zeng}
\author[1]{Jiwei Gan}
\author[1]{Qiang Hao}
\author[1,2,*]{Heping Zeng}

\affil[1]{Shanghai Key Laboratory of Modern Optical System, and Engineering Research Center of Optical Instrument and System, Ministry of Education, School of Optical Electrical and Computer Engineering, University of Shanghai for Science and Technology, Shanghai 200093, China}
\affil[2]{State Key Laboratory of Precision Spectroscopy, East China Normal University, Shanghai 200062, China}

\affil[*]{Corresponding author: hpzeng@phy.ecnu.edu.cn}

\dates{Compiled \today}

\ociscodes{(050.4865) Optical vortices; (140.3510) Lasers, fiber; (140.4050) Mode-locked lasers; (140.3615) Lasers, ytterbium.}

\doi{\url{http://dx.doi.org/10.1364/XX.XX.XXXXXX}}

\begin{abstract}
We report on a new class of mode-locked fiber laser that allows direct creation of ultrafast vector vortex beams at arbitrary positions on the higher-order Poincar\'{e} sphere. The on-demand generation of space-variant polarization patterns was realized by controlling geometric phases inside the laser resonator to map polarization to orbital angular momentum. Thanks to the ingenious cavity design, the required intracavity manipulation of the geometric phase imposed no disturbance on the passively mode-locked operation, thus demonstrating robust and flexible switching of vectorial modes with a 8.5-ps pulse duration. Analytical expressions were deduced to model the generated cylindrically-symmetric polarization profiles, and agreed exceedingly well with experimental observations. The presented fiber laser would constitute a compact light source for producing ultrafast pulses in high-purity structured modes, which may find broad applications in classical and quantum optics.
\end{abstract}

\setboolean{displaycopyright}{true}

\begin{document}

\maketitle

Vector vortex beams (VVBs) exhibit complex structuration with polarization and phase singularities in the transverse plane \cite{Maurer2007}. Among them, cylindrical vector beams (CVBs) have space-variant polarizations with cylindrical symmetry, such as radial, azimuthal and spiralling patterns \cite{Zhan2009}. These beams can be generally expressed as superpositions between two vortex beams with opposite topological charges and opposite uniform circular polarizations \cite{Padgett1999}. In analogy to the standard Poincar\'{e} sphere for geometric representation of uniform polarizations, CVBs are formally illustrated on the so-called higher-order Poincar\'{e} (HOP) sphere \cite{Milione2011} with general orthogonal bases that incorporate both spin angular momentum (SAM) and orbital angular momentum (OAM).  So far, VVBs have spawned numerous applications in modern optics (\cite{Chen2018} and references therein), among which are laser material processing, particle acceleration, optical tweezing, shaper focusing and enhanced microscopy. Additionally, unique properties such as intrinsically unbounded dimensionality, rotational invariance, and inseparability between SAM and OAM degrees of freedom make them useful in quantum information science \cite{Holleczek2011}. 

Typically, VVBs are obtained by mode transformation from uniformly polarized Gaussian beams based on spatial light modulators \cite{Wang2007}, q-plates \cite{Marrucci2006},  subwavelength gratings \cite{Bomzon2002}, metasurfaces \cite{Liu2014}, and so on. In parallel, direct generation of VVBs at the source has raised a large interest due to favorable features like compactness, high efficiency and high purity, which fuels the development of various novel light sources ranging from solid-state lasers \cite{Thirugnanasambandam2011, Ahmed2011, Kim2011, Ngcobo2013}, fiber lasers \cite{Fridman2008, Ueda2006, Lin2010, Liu2016, CarrionHigueras2017, Zhou2016, Mao2017} to integrated on-chip lasers \cite{Cai2012, Schulz2013, Miao2016}. In particular, a novel kind of laser has recently been implemented based on intracavity Pancharatnam-Berry phase control \cite{Naidoo2016}, enabling on-demand generation of all states on the HOP sphere in contrast to previous works. This technique was later adopted in a continuous-wave fiber laser to deliver switchable vectorial modes \cite{Huang2017}. However, to the best of our knowledge, no laser has so far been able to create on demand arbitrary HOP sphere beam in ultrafast pulsed excitation. In principle, the geometric phase manipulation in the aforementioned technique imposes negligible insertion loss into the laser cavity \cite{Forbes2017}, thus making it feasible to achieve ultrafast outputs based on slight modification of conventional mode-locked fiber lasers.

\begin{figure*}[t!]
\centering
\includegraphics[width=1.98\columnwidth]{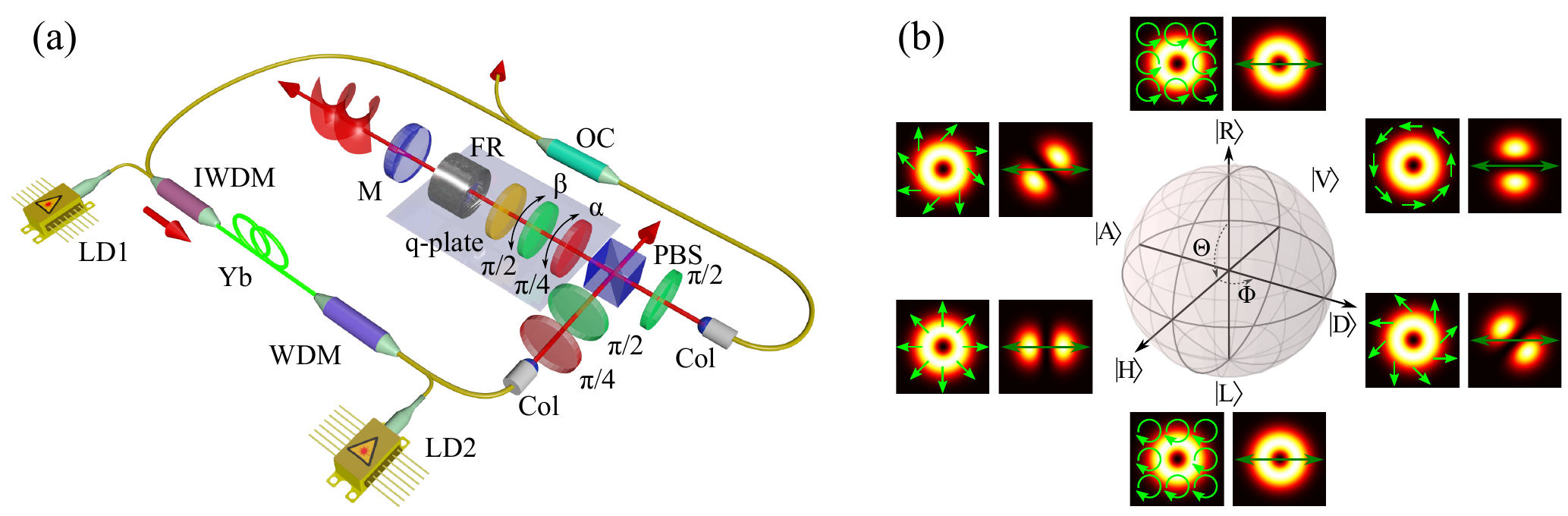}
\caption{(a). Artistic illustration of an ytterbium-doped mode-locked fiber laser for controlled generation of arbitrary vector vortex beams in an ultrafast temporal profile. The on-demand vectorial patterns are realized by judiciously rotating a quarter-wave plate and half-wave plate with angles of $\alpha$ and $\beta$, respectively. LD: Laser diode; Yb: ytterbium-doped gain fiber; WDM: wavelength division multiplexer; IWDM: WDM integrated with an isolator; FR: Faraday rotator; M: partially transmitted dielectric mirror; PBS: polarization beam splitter; OC: output coupler; Col: collimator. (b) Vector vortex beams are represented on the HOP sphere with their corresponding intensity profiles and polarization patterns. Those beams along the equator can be differentiated by the transmitted intensity distribution after a horizontally-positioned linear polarizer as denoted by the double-ended arrows. Definitions of these states are given in the text.}
\label{fig1}
\end{figure*}

In this letter, we report a mode-locked ytterbium-doped fiber laser that allows on-demand generation of ultrafast pulses in any vector vortex mode on the higher-order Poincar\'{e} sphere. In particular, scalar vortex beams on the poles and cylindrical vector beams at the equator can be produced and switched at the source by simply controlling geometric phases inside a laser cavity with a q-plate and wave plates. Moreover, the laser ring resonator is designed such that variation of output vectorial modes imposes no effect on the status of passively mode-locked operation, thus resulting in free of optical alignment and maintaining all the other properties of ultrafast pulses except for the polarization pattern. Therefore, the presented novel kind of fiber laser may permit us to engineer the fully structuration of laser fields in spatial and temporal degrees of freedom. Benefited from advanced integration technology of fiber devices, compact and turnkey light sources could be expected for directly delivering complex but well-controlled structured light in ultrashort pulses, which would facilitate many subsequent applications.

Figure \ref{fig1}(a) shows the experimental setup for the proposed mode-locked fiber laser that can deliver customized vectorial beams at the source. The configuration of the laser cavity consists of hybrid fiber and free-space sections, which is widely used in conventional fiber lasers featuring flexible cavity dispersion management and passive mode-locking operation \cite{Huang2011}. Here two laser diodes at 976 nm are used to bidirectionally pump a 1.5-meter-long ytterbium-doped gain fiber (Nufern SM-YSF-HI) through two 980/1040 nm wavelength division multiplexers (WDMs). One WDM is integrated with an isolator to form a unidirectional ring laser cavity, which favors more stable and robust laser operation. Then the fluorescence induced by the excited gain medium is steered into the free space by a home-made collimator. The collimator includes an aspheric lens with an effective focal length of 7.5 mm (Thorlabs, A375TM-B), which leads to an output beam diameter about 1.2 mm. It is worth noting that the fiber end is cleaved at an inclined angle of 8° to eliminate back reflections. After propagation in the free space, the light is collimated back into a fiber optical coupler. The 1\% tapping portion of the coupler is used for monitoring purpose, while the rest is circulated into the signal port of WDM for termination of one round trip. 

\begin{figure}[b!]
\centering
\includegraphics[width=0.98\columnwidth]{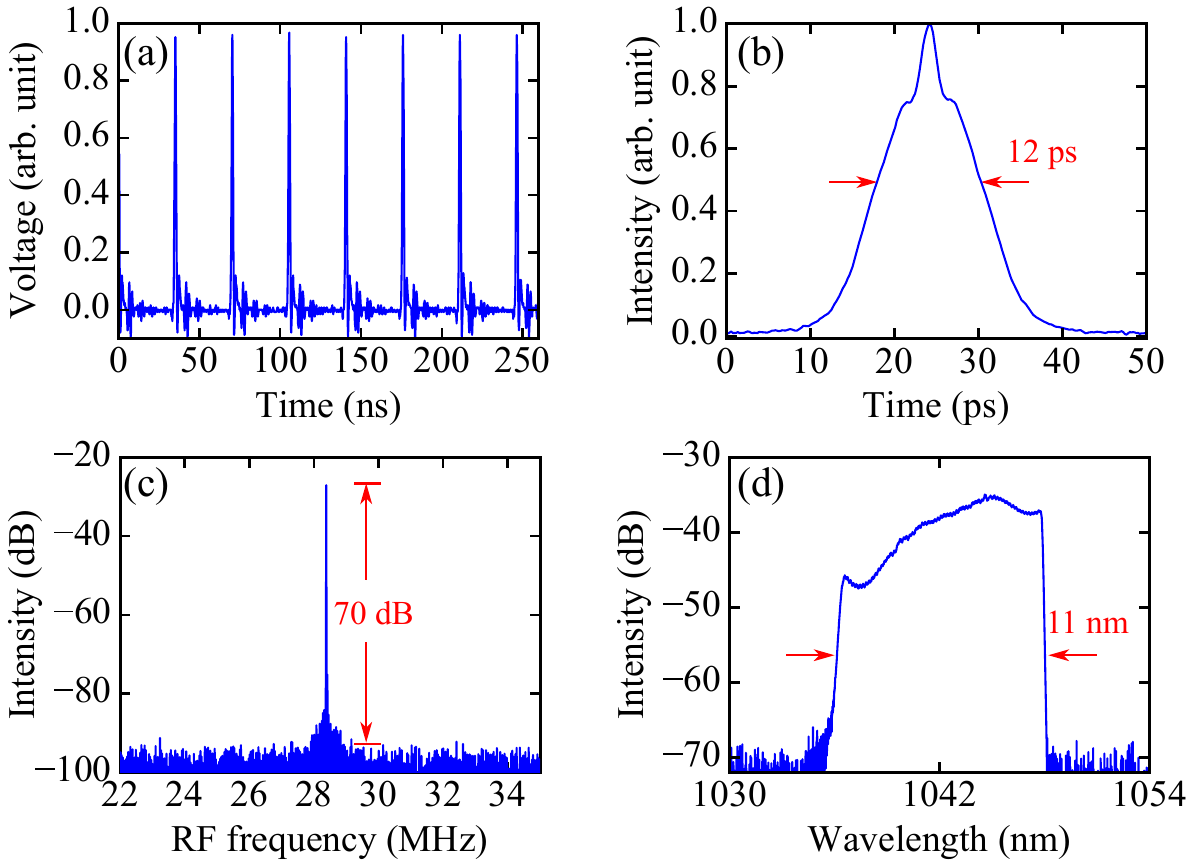}
\caption{(a) Sequence of mode-locked pulses. (b) Autocorrelation pulse profile with a FWHM duration of 12 ps. (c) Radio frequency spectrum with a signal-to-noise ratio up to 70 dB. (d) Measured optical spectrum with a FWHM bandwidth of 11 nm. These characteristics of output pulses remain identical for all the generated vector vortex beams.}
\label{fig2}
\end{figure}

The other section of the constructed laser cavity contains ingeniously arranged free-space components, which enable initiation of mode-locked pulses and tailoring of output vectorial modes. More precisely, a polarization beam splitter (PBS) and waveplates near the two collimators can act as an artificial saturable absorber thanks to the so-called nonlinear polarization rotation effect in fibers \cite{Komarov2005}. Consequently, stable mode-locking can be launched with proper angle settings of the quarter-wave plate (QWP) and the half-wave plate (HWP). Furthermore, the polarization evolution in the reflective path of the PBS (as highlighted with a shaded panel in Fig. \ref{fig1}(a)) imposes no effect on the mode-locking operation due to the presence of a Faraday rotator. It is this unique property that makes it possible to manipulate at will the space-variant polarization structuration of light in the reflective arm. Consequently, any vectorial beam on the HOP sphere can be directly output through a dielectric mirror with an intensity transmission about 4\%. 

We first characterize temporal and spectral properties of the generated pules in the experiment. To fulfill the compatibility of typical measurement devices, here the output vector beam under investigation is transformed back to the Gaussian mode by a q-plate (Thorlabs WPV10L-1064). The functionality of the q-plate will be elaborated later. With a total pump power of 350 mW, stable trains of mode-locked pulses are observed by a digital oscilloscope (Agilent MSO9404A) as shown in Fig. \ref{fig2}(a). The pulse duration is further measured to be 12 ps by an autocorrelator (APE pulseCheck) as given in Fig. \ref{fig2}(b), corresponding to actual pulse duration of 8.5 ps with assumption of a Gaussian shape. Figure \ref{fig2}(c) gives the radio frequency spectrum obtained by a spectrum analyzer (Agilent, N9000A), indicating a fundamental repetition rate of the pulses is 28.4 MHz. The achieved signal-to-noise ratio is up to 70 dB, which confirms high stability of the generated pulse sequences. The corresponding optical spectrum depicted in Fig. \ref{fig2}(d) is measured by an optical spectrum analyzer (Yokogawa AQ6370C), showing a central wavelength at 1042 nm and a FWHM bandwidth of 11 nm. The steep edges in the spectral profile is a signature for self-similar pulse evolution in a normal-dispersion cavity \cite{Prochnow2007}. Given by the significant pulse chirping, much shorter Fourier-transform-limited pulses can be obtained by internally or externally applying appropriate dispersion compensation \cite{Chong2006}. We note that the measured spectral-temporal characteristics remain identical for all vector vortex beams prepared as follows.

\begin{table}[b!]
\centering
\caption{QWP and HWP settings with corresponding output states on the HOP sphere.}
\label{table1}
  \begin{tabular}{ | l || l | l | l | l | l | l | l |}
    \hline
    \hline
    $\alpha$ & 0 & 0 & 0 & 0& $\pi$/4 & -$\pi$/4 \\ \hline
    $\beta$ &  0 & $\pi$/8 & $\pi$/4 & 3$\pi$/8 & 0 & 0 \\ \hline
    ${\left| \psi \right\rangle }$ & ${\left| A \right\rangle }$ & ${\left| V \right\rangle }$ & ${\left| D \right\rangle }$ & ${\left| H \right\rangle }$& ${\left| R\right\rangle }$& ${\left| L \right\rangle }$ \\ \hline
  \hline
  \end{tabular}
\end{table}
 
Now we turn to engineer vectorial modes of the output pulses. Generally, cylindrical vector beams (CVBs) can be expressed as linear combination of two basis states located at poles on the HOP sphere as sketched in Fig. \ref{fig1}(b). We define the two pole states as $|R\rangle  \equiv |r{\rangle} \otimes |n{\rangle}$ and $|L\rangle  \equiv |l{\rangle} \otimes | - n{\rangle}$, where $|r\rangle$ and $|l\rangle$ denote right and left circular polarizations, and $| \pm n\rangle$ are vortex states carrying $\pm n \hbar$ OAM. Consequently, radially and azimuthally polarized CVBs are given by $|H\rangle  = (|L\rangle  + |R\rangle )/\sqrt 2$ and $|V\rangle  = (|L\rangle  - |R\rangle )/\sqrt 2$, respectively. Similarly, clockwise and anticlockwise spirally polarized CVBs are given respectively by $|A\rangle  = (|L\rangle  - i|R\rangle )/\sqrt 2$ and $|D\rangle  = (|L\rangle  + i|R\rangle )/\sqrt 2$. In our experiment, the CVBs are transformed by a q-plate from generic polarization states $c_1 |r\rangle + c_2 |l\rangle$. The central idea is based on the following mapping: $|r{\rangle} \otimes |0{\rangle} \to |l{\rangle} \otimes | - 2q{\rangle} \equiv |L\rangle$ and $|l{\rangle} \otimes |0{\rangle} \to |r{\rangle} \otimes |2q{\rangle} \equiv |R\rangle$, where $q$ is the  topological charge of the q-plate \cite{Marrucci2006}. In other words, polarization control in the SAM degree of freedom provides a proxy for OAM manipulation, thus proving a simple and effective way to realize any generalized modes on the HOP sphere \cite{Naidoo2016}. As shown in Fig. \ref{fig1}(a), judicious rotation of waveplates before a q-plate with a charge of 0.5 results in an arbitrary superposition given by 
\begin{equation}
|\psi \rangle  = [\sin (\Theta /2){{\mathop{\rm e}\nolimits} ^{ - i\Phi /2}}]\left| L \right\rangle  + [\cos (\Theta /2){{\mathop{\rm e}\nolimits} ^{i\Phi /2}}]|R\rangle  \ ,
\label{eq1}
\end{equation}
where $\Theta /2  = \alpha  - \pi /4$ and $\Phi /2 = \alpha  - 2\beta  + \pi /4$ are used to simplify the expression with $\alpha$ and $\beta$ denoting rotated angles of QWP and HWP, respectively. The term $\pi/4$ in the parameter $\Phi$ originates from the rotation effect of the Faraday rotator, which manifests in the phase of the superposition coefficients as expected. Table \ref{table1} provides rotation settings of the waveplates for six specific HOP sphere beams.

\begin{figure}[t!]
\centering
\includegraphics[width=0.98\columnwidth]{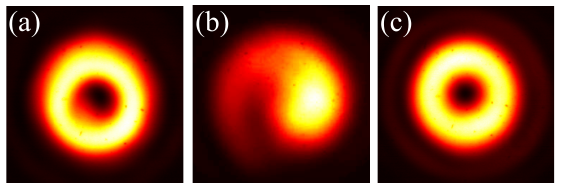}
\caption{(a) Intra-cavity generated vortex state $|R\rangle$ on the north pole of the HOP sphere. (b) Interferometric pattern between $|R\rangle$ and a Gaussian beam. The chirality of the spiral pattern indicates the sign of the carrying OAM. (c) Extra-cavity generated vortex mode which features a surrounding radial ring.}
\label{fig3}
\end{figure}

\begin{figure}[b!]
\centering
\includegraphics[width=0.95\columnwidth]{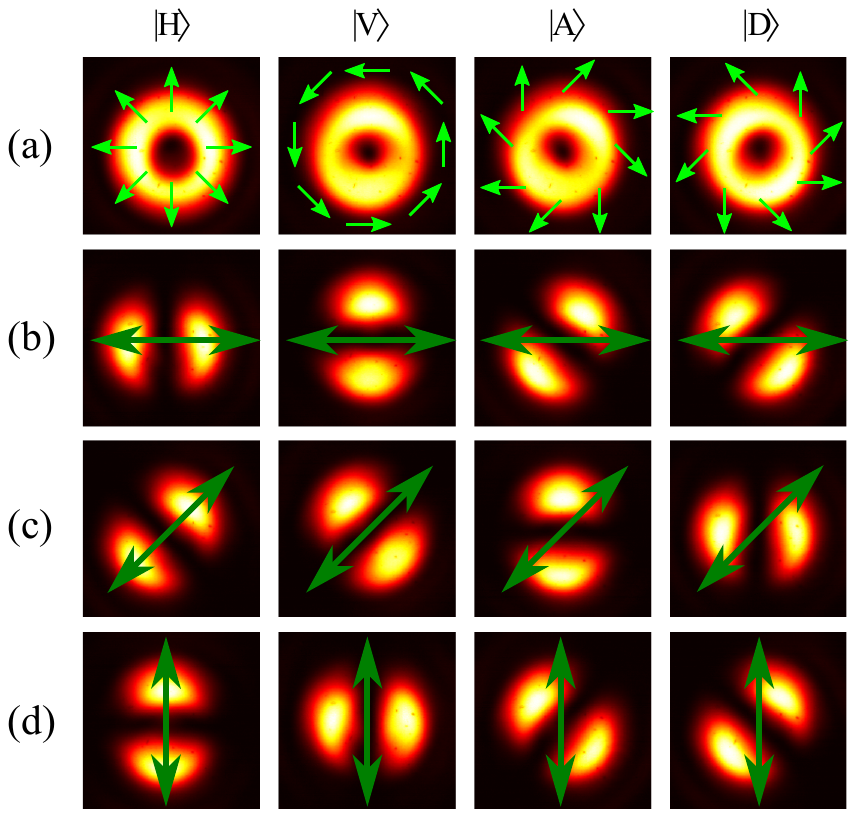}
\caption{(a) Donut-shape intensity profiles and corresponding polarization patterns are depicted for four on-demand generated cylindrical vector beams at the equator of the HOP sphere. (b)-(d) Transmitted intensity distributions after a linear polarizer with horizontal, diagonal and vertical orientations, as depicted by the superimposed double-ended arrows.}
\label{fig4}
\end{figure}

As a example, the vertically polarized light through waveplates set at $\alpha = \pi/4$ and $\beta = 0$ results in left circular polarization uniformly arranged in a Gaussian mode, $i.e.$ $|l\rangle \otimes |0\rangle$. The beam is then converted by a q-plate into an OAM beam with right circular polarization, $i.e.$ $|r\rangle \otimes |1\rangle \equiv |R\rangle$. After the output coupling mirror, the OAM beam is directly emitted from the laser as shown in an artistic fashion in Fig. \ref{fig1}(a). Subsequently, the mirror reflection flips the mode chirality as $|l\rangle \otimes |-1\rangle$. With a second pass through the q-plate and waveplates, the resultant beam is converted back to the Gaussian mode yet with a horizontal polarization. For general settings of waveplates, the ingenious cavity design can always ensure the recurrence of horizontally-polarized Gaussian spatial mode before being coupled into the fiber. Compared with the seminal work on intracavity geometric phase control \cite{Naidoo2016}, our presented scheme has two additional features. First, the use of Faraday rotator enables a line-shaped instead of V-shaped cavity arrangement, which favors a nearly perfect mode overlap between forward- and backward-propagating beams. Second, the variation of vectorial modes is realized by rotating two waveplates while retaining the q-plate at an optimal position. As a result, immobility of optical elements with space-variant properties can avoid the slight mode deformation due to imperfect alignment between the device center and rotation axis. 

The generated vector vortex beams are captured by a high-resolution CCD camera (Coherent LaserCam-HR). Figure \ref{fig3}(a) shows the intra-cavity generated state $|R\rangle$ on the north pole of the HOP sphere, which exhibits no concentric ring in contrast to the extra-cavity selected vortex beam given in Fig. \ref{fig3}(c). In theory, the mode with radial rings is generally described by a Hypergeometric-Gaussian function \cite{Sephton2016}. The high modal purity in a desirable Laguerre-Gaussian mode by the intra-cavity technique is the result of passive mode selection of laser resonator in combination with a variable aperture (not shown in Fig. \ref{fig1}(a)). Note that a Gaussian laser beam can simultaneously be obtained from one port of the PBS as shown in Fig. \ref{fig1}(a). The Gaussian beam provides a reference to obtain spiral interference patterns owing to identical spectral-temporal properties with the vortex beam. The spiral chirality can be used to distinguish the sign of the generated vortex beams $|R\rangle$ and $|L\rangle$. 

Beyond the scalar vortex beams, the presented fiber laser can generate arbitrary state on the HOP sphere. In particular,  Fig \ref{fig4}(a) gives the experimental results for four representative equatorial beams on the HOP sphere, which are superimposed with corresponding cylindrically-arranged polarization patterns. These polarization distributions are verified by observing transmitted intensity profiles after an oriented linear polarizer. The relevant mechanism lies in the non-separable nature of the vector mode between polarization and spatial subspaces. For instance, when the waveplates are set in angles of $\alpha = 0$ and $\beta = 3\pi/8$, the donut-shaped vector beam splits into two lobes that align with the orientation of the linear polarizer as shown in the column $|H\rangle$ in Fig. \ref{fig4}, thus presenting a pure radially-arranged polarization pattern. As the HWP sets at $\beta = \pi/8$, the annular beam turns to be in azimuthal polarization mode, which is hallmarked by the two-lobed structure being perpendicular to the orientation of the linear polarizer, as shown in the column $|V\rangle$ in Fig. \ref{fig4}. Similarly, spirally vectorial modes $|D\rangle$ and $|A\rangle$ can also be generated by user-defined angles of waveplates at the laser source. The experimentally generated VVBs agree well with the theoretical modes according to Eq. \ref{eq1} as shown in Fig. \ref{fig1}(b). It is worth noting that during the variation of output vector modes, the required manipulation of intracavity geometric phase imposes negligible disturbance on stably-running status of the mode-locked fiber laser, which would facilitate applications requiring continuous mode switching. Additionally, the presented technique can be extended to directly generate higher-order vector vortex beams by simply augmenting the topology charge of the q-plate \cite{Naidoo2016}.

In conclusion, we reported a novel class of mode-locked fiber laser that is able to deliver ultrafast pulses in any vectorial vortex modes on the HOP sphere. Compared to external approaches, the intracavity mode selection results in purer vector beams thanks to the passive mode filtering in a laser resonator \cite{Naidoo2016, Sephton2016}. Furthermore, the ingenious cavity configuration can be customized for other existing mode-locked fiber lasers \cite{Fermann2013}, enabling a flexible tailoring of the structured light in spatiotemporal and spectral domains. Benefited from the fiber integration technology, compact and turnkey pulsed laser sources could be envisioned to deliver vector beams with continuous switching capability and high mode purity, which may offer advantages in many subsequent applications such as microscopy, materials processing, and optical communication.

\section*{Funding.} Program for Professor of Special Appointment (Eastern Scholar) at Shanghai Institutions of Higher Learning, National Natural Science Foundation of China (11727812).

\section*{Acknowledgement.} The authors thank Dr. Darryl Naidoo for insightful discussions.

\end{document}